\documentclass[lettersize,journal]{IEEEtran}
\usepackage{amsmath,amsfonts}
\usepackage{amssymb}
\usepackage{array}
\usepackage{bm}
\usepackage[caption=false,font=footnotesize,labelfont=rm,textfont=rm]{subfig} 
\usepackage{textcomp}
\usepackage{stfloats}
\usepackage{url}
\usepackage{verbatim}
\usepackage{graphicx}
\usepackage{cite}
\usepackage[colorlinks,linkcolor=black,anchorcolor=black,citecolor=black]{hyperref} 

\usepackage{amsthm}
\usepackage{mathrsfs}
\usepackage[linesnumbered,ruled,vlined]{algorithm2e}

\begin{document}

\title{Composite Nonlinear Trajectory Tracking Control of Co-Driving Vehicles Using Self-Triggered Adaptive Dynamic Programming}

\author{Chuan Hu, Sicheng Ge, Yingkui Shi, Weinan Gao,~\IEEEmembership{Senior Member,~IEEE}, Wenfeng Guo and Xi Zhang*,~\IEEEmembership{Senior Member,~IEEE}
\thanks{This work was supported in part by the start-up fund of Shanghai Jiao Tong University, and in part by the Outstanding Youth Science Foundation Project (Overseas) of National Natural Science Foundation of China 24Z990200855. (\textit{Corresponding author: Xi Zhang.})}

\thanks{Chuan Hu and Xi Zhang are with the Intelligent Vehicle Institute,
School of Mechanical Engineering, Shanghai Jiao Tong University, Shanghai 200240, China. (e-mail: chuan.hu@sjtu.edu.cn; braver1980@sjtu.edu.cn).}

\thanks{Sicheng Ge and Yingkui Shi are with the School of Mechanical
Engineering, Shanghai Jiao Tong University, Shanghai 200240, China. (e-mail: hermit27@sjtu.edu.cn; 7120724@sjtu.edu.cn).}

\thanks{Weinan Gao is with the State Key Laboratory of Synthetical Automation for Process Industries, Northeastern University, Shenyang 110819, China. (e-mail: gaown@mail.neu.edu.cn).}

\thanks{Wenfeng Guo is with the School of Vehicle and Mobility, Tsinghua University, Beijing 100190, China (e-mail: gwf0330@163.com).}

}

    \markboth{Submitted to IEEE Transactions on Consumer Electronics}%
{Shell \MakeLowercase{\textit{et al.}}: A Sample Article Using IEEEtran.cls for IEEE Journals}


\maketitle

\begin{abstract}
This article presents a composite nonlinear feedback (CNF) control method using self-triggered (ST) adaptive dynamic programming (ADP) algorithm in a human-machine shared steering framework. For the overall system dynamics, a two-degrees-of-freedom (2-DOF) vehicle model is established and a two-point preview driver model is adopted. A dynamic authority allocation strategy based on cooperation level is proposed to combine the steering input of the human driver and the automatic controller. To make further improvements in the controller design, three main contributions are put forward. Firstly, the CNF controller is designed for trajectory tracking control with refined transient performance. Besides, the self-triggered rule is applied such that the system will update in discrete times to save computing resources and increase efficiency. Moreover, by introducing the data-based ADP algorithm, the optimal control problem can be solved through iteration using system input and output information, reducing the need for accurate knowledge of system dynamics. The effectiveness of the proposed control method is validated through Carsim-Simulink co-simulations in diverse driving scenarios. 
\end{abstract}


\section{Introduction}
\IEEEPARstart{W}{ith} the rapid progress in automation technologies and computer science, intelligent vehicles, as one of the consumer-centric products, have made significant advances and attracted much attention in recent years. Among the main research areas, automatic driving system serves as an important component which can assist the driver in controlling the vehicle, enhancing driving experience, safety and convenience. For an single vehicle, the steering control system has wide applications in maintaining lateral stability in lane keeping. A variety of research works focus on the design of vehicle lateral control strategy, including lateral path control \cite{ref1}, collision avoiding system \cite{ref2} and lane keeping system \cite{ref3}. To some extent, these studies are made based on fully autonomous vehicles, with the role of human driver being excluded and kept out-of-the-loop. However, fully autonomous driving is difficult to implement in the short term due to technological limitations and safety concerns. Human driver's insufficient attention during the driving process may result in fatal accidents when system failure occurs \cite{ref4}. Meanwhile, abrupt changes of system control input or sudden transition from fully autonomous driving to manual driving may also cause safety problems. Faced with these challenges, the human-machine shared control framework is developed where human drivers and automatic controllers cooperate interactively to accomplish different driving tasks \cite{ref5,ref6,ref7}. Schwarting \textit{et al.} \cite{ref8} propose a shared control framework for nonlinear model predictive control considering the uncertainty of road boundaries and other vehicles. In \cite{ref9,ref10,ref11}, authority allocation strategy based on driver status or road condition is implemented to integrate control inputs. Other studies also focus on driver modeling \cite{ref12}, driver intention recognition \cite{ref13} and cooperation level \cite{ref14}, aiming to realize human-vehicle collaboration. It is crucial that the objective of the shared control framework is to effectively reduce human-machine conflicts and the driver's workload while ensuring safety. 

In terms of automatic controller design, existing researches on steering control generally adopt traditional methods including proportional-integral-derivative control (PID) \cite{ref15} and model-based control algorithms such as linear quadratic regulator (LQR) and model predictive control (MPC) \cite{ref16,ref17,ref18}. However, traditional model-based control algorithms rely heavily on the prior knowledge of the system model, which might be difficult to acquire or influenced by uncertain dynamic changes of internal states. On the other hand, the Hamilton–Jacobi–Bellman (HJB) equation included in optimal control problems is often difficult to solve analytically due to its high complexity. Therefore, a large number of algorithms based on iterative learning have been developed. For instance, the adaptive dynamic programming (ADP) algorithm is an effective method to find approximate solutions for optimal control \cite{ref19,ref20}. ADP is a model-free method that learns from measurable data of the system, which is more flexible. A data-driven shared steering control method using adaptive dynamic programming is proposed in \cite{ref21}. In \cite{ref22}, ADP is used to obtain the control policy for vehicle lateral control under fully autonomous driving conditions. For nonlinear systems, \cite{ref23} proposed a critic-only ADP inspired by neural network to address tracking control problems.

Although the appearance of ADP algorithm helps to solve the control problem more easily, the control performance still has room for optimization. When driving on roads with constantly changing or large curvature, it is usually hard for vehicles to maintain steady states. Consequently, the transient performance of the system needs to be taken into consideration. The composite nonlinear feedback (CNF) control is one of the strategies that can improve the transient response. CNF control utilizes a nonlinear feedback part that can adapt the damping ratio to reduce the overshoot \cite{ref24,ref25,ref26}. In addition, existing common control methods are mostly time-triggered, meaning that the system collects data and updates control input at constant small time intervals, which is a waste of computing and communication resources. Recently, the event-driven control has been proposed, including the event-triggered control (ETC) and the self-triggered control (STC). In ETC, the system state is still measured in a time-triggered manner, but the control input is updated only when a specified trigger condition is violated \cite{ref27,ref28,ref29}. In STC, both the control input and the next triggering moment are determined and updated at the current triggering moment \cite{ref30,ref31}. It also can be seen from \cite{ref27,ref28,ref29,ref30} that the ADP algorithm also works well when ETC or STC is introduced. Generally speaking, ETC collects data continuously but updates control policy in discrete times, while STC processes all the tasks in discrete times. To this point, STC is active and might be a good choice.

Motivated by the above, this article aims to apply the data-based ADP control algorithm under human-machine co-driving process. Nevertheless, the control strategy itself can be optimized since ADP is mainly introduced as a model-free solver. We intend to improve the transient control performance and save computing resources. A CNF control method using ST-ADP algorithm in a human-machine shared steering control framework is presented. Fig. \ref{Fig:Scheme} depicts the general framework of the whole system. The main contributions are listed in the following three aspects:
\begin{figure}[!t]
    \centering
    \includegraphics[width=\linewidth]{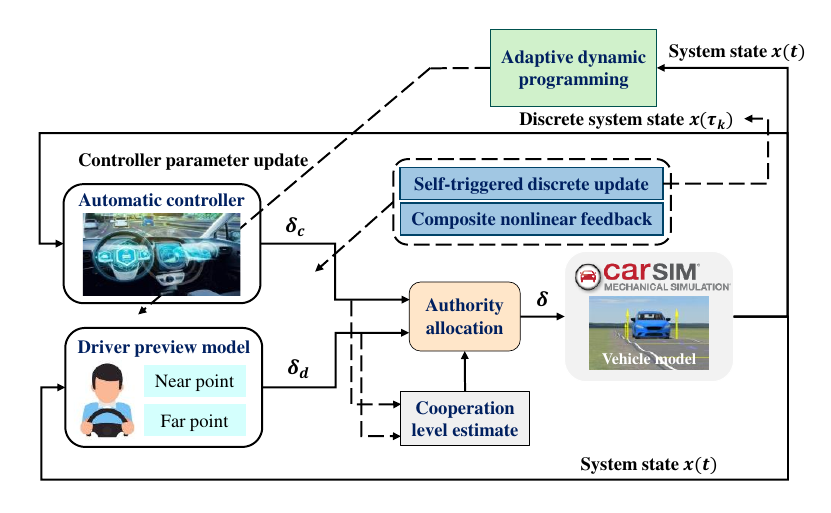}
    \caption{General framework of the co-driving system.}
    \label{Fig:Scheme}
\end{figure}

\begin{enumerate}
    \item A dynamic authority allocation strategy based on cooperation level is proposed and the system will be tested in the shared control framework. The CNF control is applied to improve the transient control performance.
    
    \item A data-based ADP algorithm is utilized to solve the optimal steering control problem by iterations without the need for accurate system model, which shows better adaptability when faced with uncertainties.

    \item To save computing and communication resources, the ST mechanism will operate after the control policy is obtained. Thus, the system can update in discrete times while realizing the optimal control.
\end{enumerate}

The rest of this article is organized as follows. Section II constructs the mathematical model for the vehicle and the human driver. An authority allocation strategy is introduced to integrate the control inputs. Section III explains the design of the automatic controller. The ST-ADP algorithm can solve the optimal problem in an iterative way and CNF control can improve transient performance. The stability of the controller is proved. Section IV verifies the effectiveness of the proposed mechanism through Carsim-Simulink simulations. Conclusions of the paper are made in Section V.

\textit{Notations:} $\mathbb{R}^n$ and $\mathbb{R}^{m\times n}$ denote the real $n$-dimensional vector and the real $m\times n$ matrix. $I_n$ denotes the $n\times n$ identity matrix. $\otimes$ is the Kronecker product. $\|\cdot\|$ is the Euclidean norm. For a symmetric matrix $M$, $M>0$ indicates that it is a positive definite matrix. $\lambda_{min}(A)$ and $\lambda_{max}(A)$ denote the minimum and maximum eigenvalue of matrix $A$. $vec(A)=[a_1^T,a_2^T,...,a_m^T]^T$, where $a_i\in\mathbb{R}^n$ are the columns of $A\in\mathbb{R}^{n\times m}$. Operators are introduced that $vecs(P)=\lbrack p_{11},2p_{12},\ldots,2p_{1m},p_{22},2p_{23},\ldots,2p_{m-1,m},p_{mm}\rbrack^T$ and $vecv(v)=\left\lbrack{v_1^2,v_1v_2,\ldots,v_1v_n,v_2^2,\ldots,v_{n-1}v_n,v_n^2}\right\rbrack^T$, where $P\in\mathbb{R}^{m\times m}$ represents square matrices and $v\in\mathbb{R}^n$ is a column vector.

\section{Control Modeling for Human-Machine Driving System}
In this section, the vehicle dynamic model and the human driver model will be established. The control inputs from the human driver and the automatic controller will be combined through shared control strategies to indicate the mechanism of the co-driving system.

\subsection{Vehicle Lateral Dynamics}
A 2-Degrees-of-Freedom (2-DOF) vehicle model is employed in this paper for steering control considering vehicle lateral dynamics. According to \cite{ref32}, a commonly nonlinear vehicle model can be linearized under the small angle hypothesis and constant longitudinal speed. Then, the state space equation can be constructed as:
\begin{figure}[!t]
    \centering 
    \includegraphics[width=0.9\linewidth]{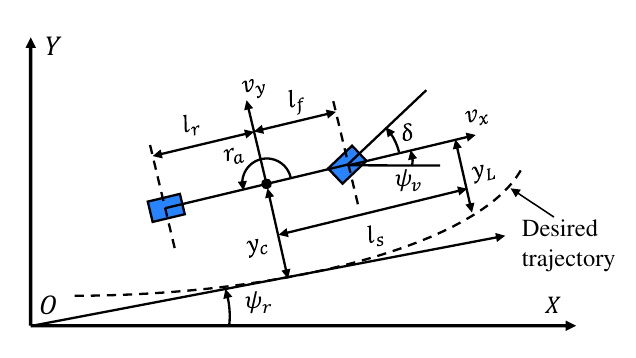}
    \caption{Vehicle dynamic model.}
    \label{Fig:VehMod}
\end{figure}
\begin{equation}
    \left\{ 
    \begin{matrix}
    {\Dot{x} = Ax + Bu + D\rho} \\
    {y = Cx}
    \end{matrix} 
    \right.
    \label{Equ:Sys}
\end{equation}
where, $x = \left\lbrack v_{y}~~r_{a}~~\psi_{L}~~y_{L} \right\rbrack^{T} $ is the vehicle states, $u=\delta$ and $y=y_c$ are the control input and output respectively, $\rho$ represents the road curvature, as illustrated in Fig. \ref{Fig:VehMod}. This vehicle model has two degrees of freedom, namely lateral deviation $y_c$ and yaw angle deviation $\psi_{L} = \psi_{v} - \psi_{r}$. $y_L$ is the lateral deviation at preview distance $l_s$. The front wheel steering angle $\delta$ is regarded as control input. $v_y$ and $r_a$ indicate the lateral velocity and yaw rate. In this model, it is assumed that the road curvature approximately remains constant in certain road sections. Besides, matrices $A, B, C$ and $D$ can be expressed as:
\begin{equation}
    \begin{split}
        &A = \begin{bmatrix}
        a_{11} & a_{12} & 0 & 0 \\
        a_{21} & a_{22} & 0 & 0 \\
        0 & 1 & 0 & 0 \\
        1 & l_{s} & v_{x} & 0
        \end{bmatrix}B = \begin{bmatrix}
        b_{1} \\
        b_{2} \\
        0 \\
        0
        \end{bmatrix}D = \begin{bmatrix}
        0 \\
        0 \\
        {- v_{x}} \\
        0
        \end{bmatrix} \\
        &C = \begin{bmatrix}
        0 & 0 & {- l_{s}} & 1
        \end{bmatrix}
    \end{split}
    \label{Equ:SysPara}
\end{equation}

The parameters in the matrices are given as follows:
\begin{equation*}
    \begin{split}
        &a_{11} = - \frac{2\left( {C_{f} + C_{r}} \right)}{mv_{x}}{~~~~a}_{12} = \frac{2\left( {C_{r}l_{r} - C_{f}l_{f}} \right)}{mv_{x}} - v_{x} \\
        &a_{21} = \frac{2\left( C_{r}l_{r} - C_{f}l_{f} \right)}{I_{z}v_{x}}{~~~a}_{22} = - \frac{2\left( C_{f}l_{f}^{2} + C_{r}l_{r}^{2} \right)}{I_{z}v_{x}} \\
        &b_{1} = \frac{2C_{f}}{m}~~~~~~~~~b_{2} = \frac{2C_{f}l_{f}}{I_{z}}
    \end{split}
\end{equation*}
where $C_f$ and $C_r$ represent the cornering stiffness of front wheel and rear wheel separately; $l_f$ and $l_r$ denote the distance from the centre of gravity to front or rear axle. $v_x$ is the constant longitudinal velocity, $I_z$ is the yaw moment of inertia, $m$ is the mass of vehicle.

\subsection{Human Driver Model}
In this paper, the human driver model is employed to simulate the driver's steering operation. A two-point preview driver model is included in this section.

\begin{figure}[!t]
    \centering
    \includegraphics[width=0.8\linewidth]{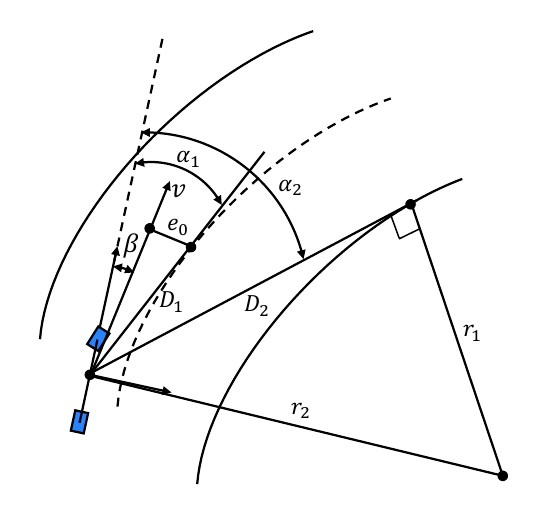}
    \caption{Human driver steering principle.}
    \label{Fig:DriMod}
\end{figure}

The visual input of this model consists of near point and far point. These two regions, whose features are denoted by the near preview angle $\alpha_1$ and the far preview angle $\alpha_2$, can represent the compensatory and anticipatory driving characters respectively, as shown in Fig. \ref{Fig:DriMod}. $\alpha_1$ and $\alpha_2$ can be obtained through the following geometric relations:
\begin{equation}
    \left\{ \begin{matrix}
    {\alpha_1=e_0/D_1+\beta} \\
    {\alpha_2=D_2/r_2+\beta}
    \end{matrix} \right.
\end{equation}
where $x_1O_1y_1$ and $x_2O_2y_2$ denote the global and local coordinate system separately. $v$ is the vehicle speed and $\beta$ is the deviation angle with respect to the direction of speed. $D_1$ and $D_2$ denote the distance from the vehicle to the near point and the far point. $r_1$ and $r_2$ denote the the radius of curvature at the far point and the radius of turning circle. $e_0$ indicates the minimum distance from the near preview point to the direction of vehicle speed.

The composition of the human driver model is illustrated in Fig. \ref{Fig:DriCom}. In this model, the human driver receives input from the two preview angles and achieves the lane-keeping process by adjusting his/her steering angle. Transfer functions $G_1(s)$ and $G_2(s)$ simulate the compensatory and anticipatory behaviours:
\begin{figure}[!t]
    \centering
    \includegraphics[width=0.8\linewidth]{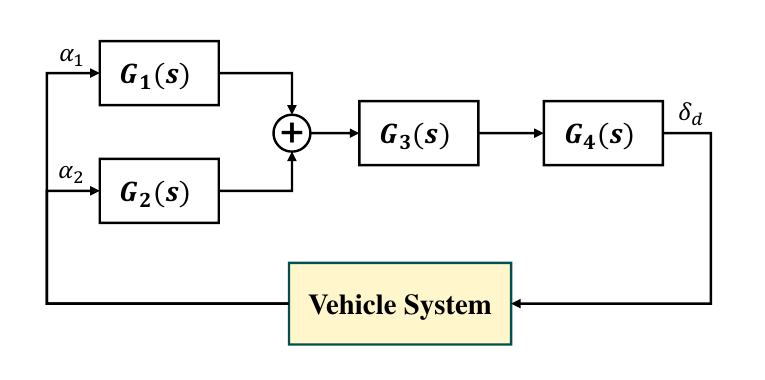}
    \caption{Composition of human driver model.}
    \label{Fig:DriCom}
\end{figure}
\begin{equation}
    \begin{cases}
    G_1(s)=\displaystyle\frac{K_1}{v}\frac{T_1s+1}{T_2s+1} \\
    G_2(s)=K_2
    \end{cases}
    \label{Equ:DriPara1}
\end{equation}
where $K_1$ and $K_2$ represent the compensatory and anticipatory control gain; $T_1$ and $T_2$ are lead and lag time constants. Besides, $G_3(s)$ represents the neuromuscular response delay and $G_4(s)$ represents the relationship between driver output and steering angle:
\begin{equation}
    \begin{cases}
    G_3(s)=\displaystyle\frac{1}{T_3s+1} \\
    G_4(s)=K_3
    \end{cases}
\end{equation}
where $T_3$ is the neuromuscular lag time constant and $K_3$ is a gain factor. With the output from the vehicle system, the driver's steering input can be derived through this model.

\subsection{Shared Control Strategy}
In the human-machine co-driving system, the control inputs are generated from both the human driver and the automatic controller. They are combined and applied to the vehicle model through an authority allocation strategy:
\begin{equation}
    u=\delta=(1-\sigma)\delta_d+\sigma\delta_c
\end{equation}
where $\sigma,1-\sigma\in[0,1]$ denote the control authorities; $\delta_d$ and $\delta_c$ denote the steering control input of the human driver and the automatic controller respectively. Similar to \cite{ref14}, an indicator for the cooperation level is introduced to dynamically adjust the control authority:
\begin{equation}
    CI(t) = \int_{t-\Delta t}^{t} \delta_d(\tau)\delta_c(\tau)d\tau
    \label{Equ:Coop}
\end{equation}

Note that this cooperativeness indicator calculates the integral of the product of two control inputs in a certain time interval. When both the driver and the automatic controller have the same objective, their corresponding steering input will be in the same direction, leading to positive cooperation indices. Otherwise, opposite steering directions will lead to negative cooperation indices, indicating a non-cooperative status. When CI takes negative value, the human driver will have a dominate control authority over the automation to avoid conflicts and perform better in emergency maneuvers. If CI is positive, the automatic controller takes more control of the vehicle to relief driver's burden. The relation between CI and the control authority is described as follows:
\begin{equation}
    \sigma = 0.5+\kappa CI(t)
    \label{Equ:Auth}
\end{equation}
where $\kappa$ is a positive gain depending on different driving scenarios. As a result, the control authority of the automatic system increases in well-cooperated conditions. We will propose the design of the automatic control system in the following chapters.

\section{Composite Nonlinear Control Using Self-Triggered Adaptive Dynamic Programming}
In this section, the trajectory tracking control method for the automatic controller will be developed. This designing process is independent of the human driver model, which means the steering operations generated by this controller will not be influenced by the human driver's decision. At first, the CNF technology is put forward to improve the transient response under the control framework based on LQR. Then, the self-triggered rule is applied to save computing resources. Moreover, the ADP algorithm is introduced into the control framework to solve the required parameters in an iterative way, realizing model-free control.

\subsection{Composite Nonlinear Feedback Method in LQR}
For general control problems, the goal is to design a feedback controller which can make the system output converge to zero. Considering the widely known linear quadratic regulator theory with a linear state space equation $\Dot{x}=Ax+Bu$, the control objective is to design a feedback control policy $u=-Kx$, which can minimize the performance index:
\begin{equation}
    J = {\int_{0}^{\infty}{\left( {x^{T}Qx + u^{T}Ru} \right)dt}}
    \label{Equ:CostF}
\end{equation}
where $Q=Q^T\geq0$, $R=R^T>0$ are weight matrices and $\left( A,\sqrt{Q} \right)$ is observable. If the system matrices are already known, the solution to the feedback control problem can be obtained directly by solving the algebraic Riccati equation (ARE) below:
\begin{equation}
    \begin{cases}
    {A^{T}P + PA + Q - PBR^{- 1}B^{T}P = 0} \\
    {K^{*} = R^{- 1}B^{T}P^{*}}
    \end{cases}
    \label{Equ:ARE}
\end{equation}
where $P$ is a unique symmetric positive definite solution.

In this paper, the lateral steering control problem with system (\ref{Equ:Sys}) has an extra disturbance $\rho$ which is regarded as a constant curvature in certain road sections. Optimal control solution for this modified system will satisfy the following equations \cite{ref33}:
\begin{equation}
    \begin{cases}
    {u = - Kx + L\rho} \\
    {L = U + KX}
    \end{cases}
    \label{Equ:Ctrl}
\end{equation}
where $K$ is the feedback gain obtained through (\ref{Equ:ARE}) based on LQR theory. Parameters $U$ and $X$ are calculated by:
\begin{equation}
    \begin{cases}
    {0 = AX + BU + D} \\
    {0 = CX}
    \end{cases}
    \label{Equ:Para}
\end{equation}
where $A, B, C$ and $D$ are system matrices.

Apart from the linear feedback law in LQR, a nonlinear part $u_N$ will also be added in the control policy, forming the CNF controller. Then, the feedback control law becomes:
\begin{equation}
    u = - Kx + L\rho + u_{N}
    \label{Equ:ContCNF}
\end{equation}

Some basic assumptions are needed for the CNF controller design with vehicle lateral dynamics (\ref{Equ:Sys}) according to \cite{ref34}:
\begin{enumerate}
    \item $A, B, C$ and $D$ are constant matrices.
    \item $(A,B)$ is stabilizable.
    \item $(A,C)$ is detectable.
    \item $(A,B,C)$ is invertible and has no zeros at $s=0$.
\end{enumerate}

The nonlinear feedback portion $u_N$ can be constructed as:
\begin{equation}
    u_{N} = \rho_{y}B^{T}Px_{e}
\end{equation}
where $\rho_{y} = - \phi e^{- \gamma y}$ is some continuously differentiable function of the control output $y$ with positive parameters $\phi$ and $\gamma$, $x_{e} = x - X\rho$ and $P$ is the positive definite solution in (\ref{Equ:ARE}). It is worth noticed that $u_N$ is aimed to increase the damping ratio of the system when the control output $y$ is close to zero, thus reducing the overshoot and improving the transient performance. The stability of this CNF method will be proved together with the event-triggered control design.

\subsection{Event-Triggered Control (ETC)}
Some basic principles of event-triggered control are introduced before self-triggered strategy is applied. In event-triggered algorithm, the control input $u$ is updated at certain triggering instants denoted as $\left\{ \tau_{k} \right\}_{k = 0}^{\infty}$ where $\tau_0=0$, $\tau_{k} < \tau_{k + 1} \in \mathbb{R}^{+}$ and $k \in \mathbb{N}$. These sampling triggering instants occur when the error caused by ETC exceeds the event-triggered threshold. This error and the triggering condition can be defined by:
\begin{equation}
    {e(t) = x_{k} - x(t),~}t \in \left\lbrack \tau_{k},\tau_{k + 1} \right)
\end{equation}
and
\begin{equation}
    \left\| {e(t)} \right\|^{2} > e_{T}
\end{equation}
where $x_k$ is the state vector sampled at time $\tau_k$, namely $x(\tau_k)$. $x(t)$ denotes the current state vector. $e_T$ is the event-triggered threshold. Error $e(t)$ is reset to zero at every triggering instant since state vectors are sampled. Meanwhile, control input (\ref{Equ:ContCNF}) also remains constant within the time interval $[\tau_k,\tau_{k+1})$:
\begin{equation}
    u_{k} = - Kx_{k} + L\rho + u_{N}
    \label{Equ:DiscCNF}
\end{equation}

It should be noted that under the event-triggered mechanism, although both $x_k$ and $u_k$ can update only at certain time instants, the current system state $x(t)$ and error $e(t)$ need to be collected continuously.

The stability of the system will be guaranteed under both event-triggered and self-triggered mechanisms. At first, two basic assumptions on Lipschitz continuity are given \cite{ref30}:

\textit{Assumption 1:} $f(x)=Ax$ is Lipschitz continuous on a compact set $\Omega$, so there exists $L_f>0$ such that $\left\| {f(x)} \right\| \leq L_{f}\left\| x \right\|$.

\textit{Assumption 2:} $u(x)$ is Lipschitz continuous on set $\Omega$, so there exists a constant $L_u>0$ such that $\left\| {u(x) - u\left( x_{k} \right)} \right\| \leq L_{u}\left\| {x - x_{k}} \right\|$.

The event triggered rule is given as the following theorem.

\textit{Theorem 1:} Suppose Assumptions 1 and 2 hold. Let $Q$ be the weight matrix in (\ref{Equ:CostF}), the closed-loop system described by (\ref{Equ:Sys}) and (\ref{Equ:Ctrl}) is asymptotically stable under the event-triggered condition as follows \cite{ref28,ref30}:
\begin{equation}
    \left\| {e(t)} \right\|^{2} > \frac{\left( {1 - \alpha} \right)\lambda_{min}(Q)}{\left( {\frac{1}{\alpha} - 1} \right)\lambda_{max}(Q)}\left\| x_{ek} \right\|^{2} = e_{T}
    \label{Equ:ET}
\end{equation}
where, $\alpha\in(0,1)$, $x_{ek}$ is the sample of $x_e$ at $t=\tau_k$.

\textit{Proof:} Substitute the control input (\ref{Equ:DiscCNF}) into the system state space equation (\ref{Equ:Sys}), we have:
\begin{equation}
    \Dot{x} = Ax - BK\left( {x + e} \right) + B\left( {U + KX} \right)\rho + D\rho + Bu_{N}
\end{equation}

Using (\ref{Equ:Para}), the equation can be rewritten as:
\begin{equation}
    \Dot{x} = \left( {A - BK} \right)\left( {x - X\rho} \right) + Bu_{N} - BKe
\end{equation}

Adopting $A_k=A-BK$ and $x_e=x-X\rho$, we get:
\begin{equation}
    \Dot{x_{e}} = {A_{k}x}_{e} + Bu_{N} - BKe
    \label{Equ:StaPoint}
\end{equation}

It is obvious that $x_e=0$ is a stable point in (\ref{Equ:StaPoint}). Construct a Lyapunov function that takes the following form:
\begin{equation}
    V = {x_{e}}^{T}Px_{e}
\end{equation}

Let $Q_k=Q+PBR^{-1}B^TP$, the derivative of $V$ yields:
\begin{equation}
    \begin{aligned}
        \Dot{V} 
        &= \Dot{x_e}^TPx_{e} + x_{e}^{T}P\Dot{x_{e}}\\
        &= \left( {x_{e}^{T}A_{k}^{T} + B^{T}u_{N} - e^{T}K^{T}B^{T}} \right)Px_{e}\\
        &+ x_{e}^{T}P\left( {A_{k}x}_{e} + Bu_{N} - BKe \right)\\
        &= {- x}_{e}^{T}Q_{k}x_{e} + u_{N}\left( {B^{T}Px_{e} + x_{e}^{T}PB} \right) - 2x_{e}^{T}PBKe
    \end{aligned}
    \label{Equ:dV}
\end{equation}

Note that the error $e$ can also be expressed as $e=x_{ek}-x_e$, so the first term in (\ref{Equ:dV}) becomes:
\begin{equation}
    {-x}_{e}^{T}Q_{k}x_{e} = {-x}_{ek}^{T}Q_{k}x_{ek} + {2x}_{ek}^{T}Q_{k}e - {e^{T}Q}_{k}e
    \label{Equ:fstT}
\end{equation}

The last term in (\ref{Equ:dV}) becomes:
\begin{equation}
    {-2x}_{e}^{T}PBKe = {-2x}_{ek}^{T}PBKe + 2e^{T}PBKe
    \label{Equ:lastT}
\end{equation}

Combining (\ref{Equ:fstT}) and (\ref{Equ:lastT}), it can be derived that:
\begin{equation}
    \begin{aligned}
         &{-x}_{e}^{T}Q_{k}x_{e}-2x_{e}^{T}PBKe\\
         &={-x}_{ek}^{T}Q_{k}x_{ek} + {2x}_{ek}^{T}Qe - e^{T}Qe + e^{T}PBKe
    \end{aligned}
    \label{Equ:Combine}
\end{equation}

A formula can be obtained through the average value inequality that:
\begin{equation}
    {2x}_{ek}^{T}Qe \leq \alpha x_{ek}^{T}Qx_{ek} + \frac{1}{\alpha}e^{T}Qe
\end{equation}
where $\alpha\in(0,1)$. Then, equation (\ref{Equ:Combine}) becomes:
\begin{equation}
    \begin{aligned}
         &{-x}_{e}^{T}Q_{k}x_{e}-2x_{e}^{T}PBKe\leq \left( {\alpha - 1} \right)x_{ek}^{T}Qx_{ek}\\
         &-x_{ek}^{T}PBKx_{ek}-\left( {1 - \frac{1}{\alpha}} \right)e^{T}Qe + e^{T}PBKe
    \end{aligned}
    \label{Equ:ETIn3}
\end{equation}

From the second term in (\ref{Equ:dV}), we obtain:
\begin{equation}
    \begin{aligned}
         &u_{N}\left( {B^{T}Px_{e} + x_{e}^{T}PB} \right) = 2u_{N}B^{T}P\left( {x_{ek} - e} \right)\\
         &=2\rho_{y}\left( {x_{ek}^{T}PBB^{T}Px_{ek} - x_{ek}^{T}PBB^{T}Pe}\right)
    \end{aligned}
    \label{Equ:ETIn4}
\end{equation}

Under the event-triggered condition (\ref{Equ:ET}), by taking a suitable value for $\alpha\in(0,1)$, the following inequalities hold that:
\begin{equation}
    \left( {\frac{1}{\alpha} - 1} \right)e^{T}Qe \leq \left( {1 - \alpha} \right)x_{ek}^{T}Qx_{ek}
    \label{Equ:ETIn1}
\end{equation}
\begin{equation}
    e^{T}PBKe \leq x_{ek}^{T}PBKx_{ek}
    \label{Equ:ETIn2}
\end{equation}

As a result of (\ref{Equ:ETIn1}) and (\ref{Equ:ETIn2}), both (\ref{Equ:ETIn3}) and (\ref{Equ:ETIn4}) become negative. Thus, the derivative of the Lyapunov function satisfies $\Dot{V}\leq0$. In particular, the value of $\Dot{V}$ will not be identically equal to zero due to the constant update and reset of error $e$. Consequently, the system is asymptotically stable.

\subsection{Self-Triggered Control (STC) Design}
Next, the self-triggered rule is established based on the event-triggered condition. Unlike ETC, self-triggered control calculates the next triggering instant without continuous state measurement. The stability of the system will be proved.

\textit{Theorem 2:} Consider the event-triggered condition obtained through (\ref{Equ:ET}) under the system with (\ref{Equ:Sys}) and (\ref{Equ:Ctrl}). The system is asymptotically stable if the self-triggered rule is designed as follows \cite{ref30}:
\begin{equation}
    \begin{aligned}
         &\tau_{k + 1} = \tau_{k} + \frac{1}{a + b}ln\left( 1 + \frac{a + b}{a\left\| x_{ek} \right\| + c} \right.\\
         &\left. \times \sqrt{\frac{\left( {1 - \alpha} \right)\lambda_{min}(Q)}{\left( {\frac{1}{\alpha} - 1} \right)\lambda_{max}(Q)}\left\| x_{ek} \right\|^{2}} \right)
    \end{aligned}
    \label{Equ:ST}
\end{equation}
where $a, b$ and $c$ are positive scalars.

\textit{Proof:} In (\ref{Equ:StaPoint}), let $u_{ek}=u_k-U\rho=-Kx_{ek}+u_N$ and $u_e=u-U\rho=-Kx_e+u_N$, the state space equation can be rewritten as:
\begin{equation}
    \Dot{x_{e}} = {Ax}_{e} + Bu_{ek}
\end{equation}

Using Assumptions 1 and 2, the following inequalities can be derived:
\begin{equation}
    {\left\| \Dot{x_{e}} \right\|} \leq \left\| {Ax}_{e} \right\| + \left\| {Bu_{ek}} \right\| \leq L_{f}\left\| x_{e} \right\| + \left\| {Bu_{ek}} \right\|
    \label{Equ:STIn1}
\end{equation}
\begin{equation}
    \left\| u_{ek} \right\| \leq \left\| {u_{e} - u_{ek}} \right\| + \left\| u_{e} \right\| \leq L_{u}\left\| {e(t)} \right\| + \left\| u_{e} \right\|
\end{equation}

Thus, (\ref{Equ:STIn1}) yields
\begin{equation}
    \left\| \Dot{x_{e}} \right\| \leq L_{f}\left\| x_{e} \right\| + \left\| B \right\| L_{u}\left\| {e(t)} \right\| + \left\| {Bu_{e}} \right\|
    \label{Equ:STIn1Y}
\end{equation}

Suppose $\|u_e\|$ is bounded and $\|u_e\|\leq\varphi$ with $\varphi$ being a positive constant. Let $a=L_f$, $b=\|B\|L_u$, $c=\|B\|\varphi$, we can rewrite (\ref{Equ:STIn1Y}) as:
\begin{equation}
    {\left\| \Dot{x_{e}} \right\|} \leq a\left\| x_{e} \right\| + b\left\| {e(t)} \right\| + c
\end{equation}

For every time interval $t\in[\tau_k,\tau_{k+1})$, we have:
\begin{equation}
    \left\| \Dot{e(t)} \right\| = {\left\| \Dot{x_{e}} \right\|} \leq a\left\| x_{ek} \right\| + (a + b)\left\| {e(t)} \right\| + c
\end{equation}

According to the comparing lemma \cite{ref35}, the error $\|e(t)\|$ will be restricted with respect to time:
\begin{equation}
    \left\| {e(t)} \right\| \leq \frac{a\left\| x_{ek} \right\| + c}{a + b}\left( {e^{{({a + b})}{({t - \tau_{k}})}} - 1} \right)
    \label{Equ:STerr}
\end{equation}

So, when the triggering threshold is exceeded and the control input updates at $t=\tau_{k+1}$, (\ref{Equ:STerr}) becomes:
\begin{equation}
    \frac{a\left\| x_{ek} \right\| + c}{a + b}\left( {e^{{({a + b})}{({\tau_{k + 1} - \tau_{k}})}} - 1} \right) \geq \left\| {e\left( \tau_{k + 1} \right)} \right\| > \sqrt{e_{T}}
\end{equation}

In other words:
\begin{equation}
    \tau_{k + 1} - \tau_{k} > \frac{1}{a + b}{\ln\left( {1 + \frac{a + b}{a\left\| x_{ek} \right\| + c}\sqrt{e_{T}}} \right)} > 0
    \label{Equ:STrule}
\end{equation}

Thus, the self-triggered rule (\ref{Equ:ST}) is derived.

\textit{Remark 1:} In event-triggered control, constant measurement of the system state is needed since the event-triggered condition (\ref{Equ:ET}) contains error $e(t)$, which is related to continuous time. When the event-triggered rule (\ref{Equ:ET}) is satisfied, a new control input $u_k$ is generated based on the vehicle state measured at the new triggering instant. Otherwise, the control input remains the same during the time interval $[\tau_k,\tau_{k+1})$. 

\textit{Remark 2:} In self-triggered control, the next triggering time instant is calculated in advance using only current time and vehicle states, which means periodical measurements are not necessary. Since STC collects less system state information, its triggering time interval is shorter than ETC, so STC is more conservative. In other words, STC updates the system state $x_k$ and the control input $u_k$ at discrete times, while ETC needs continuous state measurements but updates $u_k$ with longer time intervals. 

\textit{Remark 3:} It is necessary to avoid Zeno behavior in ETC or STC which means an infinite number of triggers occurring in a finite time interval. The minimum sampling time interval must be positive. Noticing that the time interval derived from (\ref{Equ:STrule}) always remains positive, Zeno behavior can be avoided.

\subsection{Adaptive Dynamic Programming Algorithm}
Generally, it is difficult to solve $P^\ast$ and the feedback gain $K^\ast$ directly from ARE (\ref{Equ:ARE}). To solve this problem, some technical algorithms have been developed. One of these algorithms solves this problem recursively as described in \cite{ref36}:  

\textit{Theorem 3:} Suppose an initial stabilizing feedback gain $K_0$ is available. Let $P_j$ be the solution of the Lyapunov equation:
\begin{equation}
    \left( A - BK_{j} \right)^{T}P_{j} + P_{j}\left( A - BK_{j} \right) + Q + K_{j}^{T}RK_{j} = 0
\end{equation}
where $K_j$ is given by:
\begin{equation}
    K_{j} = R^{- 1}B^{T}P_{j - 1}
\end{equation}

Then, the following properties hold:
\begin{enumerate}
    \item Matrix $A-BK_j$ is Hurwitz.
    \item $P^{*}\leq P_{j + 1}\leq P_{j}$ and ${\lim\limits_{j\rightarrow\infty}K_{j}} = K^{*}$, ${\lim\limits_{j\rightarrow\infty}P_{j}} = P^{*}$.
\end{enumerate}

Nevertheless, this iteration method requires the knowledge of the system matrices $A$ and $B$, whose accurate values are often unavailable. In this case, a model-free data-driven approach is introduced based on ADP. In this method, the problem can be solved even though the system matrices are unknown. Instead, it learns from the measurement of system states. Similar to \cite{ref33}, this iterative method is established in the following.

To figure out the control policy defined in (\ref{Equ:Ctrl}), approximate values for $K, U$ and $X$ need to be obtained. First, parameter $X$ can be expressed by:
\begin{equation}
    X = Y^{1} + {\sum\limits_{l = 2}^{4}{\alpha^{l}Y^{l}}}
    \label{Equ:ADPX}
\end{equation}
where $Y^{1} = 0_{4\times1}$ and for $l=2,3,4$, $\alpha^l\in\mathbb{R}$, $Y^l\in{\mathbb{R}}^4$ and $CY^l=0$. Let $x^{l} = x - Y^{l}\rho$, the system equation (\ref{Equ:Sys}) becomes:
\begin{equation}
    \begin{aligned}
        \Dot{x}^{l} &= Ax + Bu + D\rho \\
        &= A_{j}x^{l} + B\left( {K_{j}x^{l} + u} \right) + \left( D + AY^{l} \right)\rho
    \end{aligned}
\end{equation}
where $A_j=A-BK_j$.

Then, take the derivative of $\left( x^{l} \right)^{T}P_{j}x^{l}$, we yield:
\begin{equation}
    \begin{aligned}
        &\left(\Dot{x}^{l}\right)^TP_jx^l
        + \left(x^l\right)^TP_j\Dot{x}^{l} \\
        &= \left(x^l\right)^T\left( {A_{j}^{T}P_{j} + P_{j}A_{j}} \right)x^{l} + 2\left( {K_{j}x^{l} + u} \right)B^{T}P_{j}x^{l} \\
        &+ 2\rho_{i}\left( D + AY^{l} \right)^{T}P_{j}x^{l} \\
        &= - \left( x^{l} \right)^{T}\left( {Q + rK_{j}^{T}K_{j}} \right)x^{l} + 2r\left( {K_{j}x^{l} + u} \right)K_{j + 1}x^{l} \\
        &+ 2\rho\left( D + AY^{l} \right)^{T}P_{j}x^{l}
    \end{aligned}
\end{equation}
where $r=R^{1\times1}$ is a constant in weight matrix in (\ref{Equ:CostF}). Moreover, in some time interval $\delta t$:
\begin{equation}
    \begin{aligned}
        &\left({x^l\left({t+\delta t} \right)} \right)^{T}P_{j}x^{l}\left( {t + \delta t} \right) - \left( {x^{l}(t)} \right)^{T}P_{j}x^{l}(t) \\
        = &- {\int_{t}^{t + \delta t}{\left( x^{l} \right)^{T}\left( {Q + rK_{j}^{T}K_{j}} \right)x^{l}d\tau}} \\
        &+ {\int_{t}^{t + \delta t}{2r\left( {K_{j}x^{l} + u} \right)K_{j + 1}x^{l}d\tau}} \\
        &+ {\int_{t}^{t + \delta t}{2\rho\left( D + AY^{l} \right)^{T}P_{j}x^{l}d\tau}}
    \end{aligned}
    \label{Equ:ADPTI}
\end{equation}

For the sake of concise expression, the terms in (\ref{Equ:ADPTI}) can be further written in the form of Kronecker product representation:
\begin{equation}
    \begin{aligned}
        \left( x^{l} \right)^{T}\left( {Q + rK_{j}^{T}K_{j}} \right)x^{l}&= \left\lbrack  \left( x^{l} \right)^{T} \otimes\left( x^{l} \right)^{T} \right\rbrack vec\left( {Q + rK_{j}^{T}K_{j}} \right)\\
        r\left( {K_{j}x^{l} + u} \right)K_{j + 1}x^{l}&=\Big\{\left\lbrack { \left( x^{l} \right)^{T} \otimes\left( x^{l} \right)^{T}} \right\rbrack\left( {I_{4} \otimes rK_{j}^{T}} \right) \\
        &+ r\left\lbrack \left( x ^{l} \right)^{T} \otimes u \right\rbrack I_{4}\Big\} vec\left( K_{j + 1} \right)\\
        \rho\left( D + AY^{l} \right)^{T}P_{j}x^{l} &= \left\lbrack \left( x^{l} \right)^{T} \otimes \rho \right\rbrack vec\left( {\left( {D + AY^{l}} \right)^{T}P_{j}} \right) \\
        \left( {x^{l}\left( {t + \delta t} \right)} \right)^{T}&P_{j}x^{l}( {t + \delta t}) - \left( {x^{l}(t)} \right)^{T}P_{j}x^{l}(t) \\
        =\Big\lbrack vecv\left( {x^{l}\left( {t + \delta t} \right)} \right)& - vecv\left( {x^{l}(t)} \right) \Big\rbrack ^{T} vecs\left( P_{j} \right)
    \end{aligned}
\end{equation}

In time interval $\lbrack t_0,t_s\rbrack$ where $s\in\mathbb{N}$, the system state $x$, the control input $u$ and the constant road curvature $\rho$ are recorded and matrices $\delta_{x^{l}x^{l}}$, $\Gamma_{x^{l}x^{l}}$, $\Gamma_{x^{l}u}$ and $\Gamma_{x^{l}\rho}$ are defined:
\begin{equation}
    \begin{aligned}
        \delta_{x^lx^l} &= \big\lbrack vecv\left({x^l\left(t_1 \right)} \right) - vecv\left( {x^{l}\left( t_{0} \right)} \right),vecv\left( {x^{l}\left( t_{2} \right)} \right) \\
        & -vecv\left( {x^{l}\left( t_{1} \right)} \right),\ldots,vecv\left( {x^{l}\left( t_{s} \right)} \right) - vecv\left( {x^{l}\left( t_{s - 1} \right)} \right) \big\rbrack^{T} \\
        \Gamma_{x^{l}x^{l}} &= \left\lbrack {{\int_{t_{0}}^{t_{1}}{x^{l} \otimes x^{l}d\tau}},{\int_{t_{1}}^{t_{2}}{x^{l} \otimes x^{l}d\tau}},\ldots,{\int_{t_{s - 1}}^{t_{s}}{x^{l} \otimes x^{l}d\tau}}} \right\rbrack^{T} \\
        \Gamma_{x^{l}u} &= \left\lbrack {{\int_{t_{0}}^{t_{1}}{x^{l} \otimes ud\tau}},{\int_{t_{1}}^{t_{2}}{x^{l} \otimes ud\tau}},\ldots,{\int_{t_{s - 1}}^{t_{s}}{x^{l} \otimes ud\tau}}} \right\rbrack^{T} \\
        \Gamma_{x^{l}\rho} &= \left\lbrack {{\int_{t_{0}}^{t_{1}}{x^{l} \otimes \rho d\tau}},{\int_{t_{1}}^{t_{2}}{x^{l} \otimes \rho d\tau}},\ldots,{\int_{t_{s - 1}}^{t_{s}}{x^{l} \otimes \rho d\tau}}} \right\rbrack^{T}
    \end{aligned}
\end{equation}
where $0\leq t_0<t_1<\ldots<t_s$ are time instants. Therefore, the iteration process (\ref{Equ:ADPTI}) can be expressed in the following matrix form:
\begin{equation}
    \begin{bmatrix}
    {vecs\left( P_{j} \right)} \\
    {vec\left( K_{j + 1} \right)} \\
    {vec\left( {\left( {D + AY^{l}} \right)^{T}P_{j}} \right)}
    \end{bmatrix} 
    = \left( {\Theta_{j}^{T}\Theta_{j}} \right)^{- 1}\Theta_{j}^{T}\Xi_{j}
    \label{Equ:ADPIter}
\end{equation}
where 
\begin{equation}
    \begin{aligned}
        &\Theta_{j} = \left\lbrack {\delta_{x^{l}x^{l}}, - 2\Gamma_{x^{l}x^{l}}\left( {I_{4} \otimes rK_{j}^{T}} \right) - 2\Gamma_{x^{l}u}\left( {rI_{4}} \right), - 2\Gamma_{x^{l}\rho}} \right\rbrack \\
        &\Xi_{j} = - \Gamma_{x^{l}x^{l}}vec\left( {Q + rK_{j}^{T}K_{j}} \right)
    \end{aligned}
    \label{Equ:ADPMat}
\end{equation}

It should be noted that (\ref{Equ:ADPIter}) holds when matrix $\Theta_j$ has full column rank, which can be achieved through sufficient data sampling. Besides, some exploration noise $\xi$ is employed at the beginning such that $u=-K_0x+\xi$. Through the iteration learning (\ref{Equ:ADPIter}), the control feedback gain $K$ and the symmetric positive-definite matrix $P$ can be solved. Furthermore, parameters $U$ and $X$ are also necessary in the control policy. Note that $D$ can be computed when $l=1$ in (\ref{Equ:ADPIter}) since $Y^1=0$. Subsequently, $AY^l$ can be determined when $l=2,3,4$ using the knowledge of $D$ and $B$ can be estimated by $rP_{j}^{- 1}K_{j + 1}^{T}$. Thus, $U$ and $\alpha^l(l=2,3,4)$ can be solved through the rewritten form of (\ref{Equ:Para}):
\begin{equation}
    {\sum\limits_{l = 2}^{4}{\alpha^{l}AY^{l}}} + BU + D = 0
    \label{Equ:ADPU}
\end{equation}
Consequently, $X$ can be obtained through (\ref{Equ:ADPX}). The whole process is summarized in Algorithm \ref{Algorithm1}.

\begin{algorithm}
\SetAlgoLined 
\caption{Model-free ADP control}\label{Algorithm1}
\KwData{Set the weight matrix $Q=diag(100,100,100,100)$ and $R=100$. Select a proper $K_0$ that stabilizes the initial system. Let $Y^1=0_{4\times1}$ and calculate $Y^l$ satisfying $Y^l\in{\mathbb{R}}^4$ and $CY^l=0$ for $l=2,3,4$.}
$j\leftarrow 0$\;
\While{$\Theta_j$ doesn't have full column rank}{
    Apply the control input $u=-K_0x+\xi$\;
    \For{$l=1,2,3,4$}{
        Compute $\Theta_{j}$ and $\Xi_{j}$ from (\ref{Equ:ADPMat})\;
    }
}
Solve $P_j$ and $K_{j+1}$ from (\ref{Equ:ADPIter})\;
\While{$\left| {P_j - P_{j - 1}} \right| > err$}{
    $j\leftarrow j+1$\;
    Update $P_j$ and $K_{j+1}$ from (\ref{Equ:ADPIter})\;
}
\For{$l=1,2,3,4$}{
    Apply the optimal value of $P$ and $K$\;
    Compute $D$ for $l=1$\;
    Compute $AY^l$ for $l=2,3,4$\;
}
Estimate $B$ by $rP_{j}^{- 1}K_{j + 1}^{T}$\;
Compute $U$ and $X$ through (\ref{Equ:ADPU}) and get $L$ in (\ref{Equ:Ctrl})\;
For the nonlinear part in CNF control, we get $u_{N} = \rho_{y}B^{T}Px_{e}$ where $\rho_{y} = - \phi e^{- \gamma y}$ and $x_{e} = x - X\rho$\;
\KwResult{The optimal control policy is derived: $u =-Kx+L\rho+u_N$.}
\end{algorithm}

\textit{Remark 4:} The method of adaptive dynamic programming is applied to solve the feedback gain $K$ and matrix $P$ offline through system states measured continuously over time instead of knowing the system matrices. The nonlinear control law can also be determined this way since the required parameters are already included in the iteration. After the optimal control policy is established, the self-triggered rule starts working and the system will update at discrete times. If the system dynamics is not fixed and the parameters change, a new ADP process will take place to update the control policy.

\textit{Remark 5:} For a data-based approach, the system input and output information will definitely influence the system states and furthermore, the quality of the data collected as well as the calculation results. However, we do not intend to discuss the impact of data quality on the computational results in detail in this paper. The convergence of parameters $P$ and $K$ can be ensured providing that a proper $K_0$ is selected to stabilize the system in the algorithm.

\section{Simulation Results and Analysis}
In this section, Carsim-Simulink simulations are conducted to verify the  effectiveness of the proposed shared steering control method. Firstly, the vehicle model is constructed in Carsim and imported to Simulink in the form of S-Function. Besides, a two-point preview driver model is also set up in Simulink. Table \ref{table_1} lists the parameters of the vehicle model and the driver model. The parameters in the left column can be referred to the vehicle dynamics in (\ref{Equ:SysPara}), while the right column contains time constants and gains that can be referred to (\ref{Equ:DriPara1})-(\ref{Equ:Auth}). $L=l_f+l_r$ represents the distance between front axle and rear axle.
\begin{table}[!t]
    \renewcommand{\arraystretch}{1.25}
    \centering
    \caption{Parameters for System Simulation}
    \scalebox{1.15}{
        \begin{tabular}{cc|cc}
        \hline \hline
        Parameters & Value & Parameters & Value\\
        \hline
        $m$ & 1370$\mathrm{kg}$ & $K_1$ & 15 \\
        $l_f$ & 1.11$\mathrm{m}$ & $K_2$ & 3.4 \\
        $l_r$ & 1.756$\mathrm{m}$ & $K_3$ & 1/12 \\
        $L$ & 2.866$\mathrm{m}$ & $T_1$ & 3$\mathrm{s}$ \\
        $C_f$ & 56300$\mathrm{N/rad}$ & $T_2$ & 1$\mathrm{s}$ \\
        $C_r$ & 47250$\mathrm{N/rad}$ & $T_3$ & 0.1$\mathrm{s}$ \\
        $I_z$ & 2315$\mathrm{kg\cdot m^2}$ & $\Delta t$ & 5$\mathrm{s}$ \\
        $l_s$ & 5$\mathrm{m}$ & $\kappa_1$ & 5 \\
        $v_x$ & 15$\mathrm{m/s}$ & $\kappa_2$ & 5 \\
        \hline
        \hline
        \end{tabular}
    }
    \label{table_1}
\end{table}

\begin{figure}[!t]
    \centering
    \includegraphics[width=0.9\linewidth]{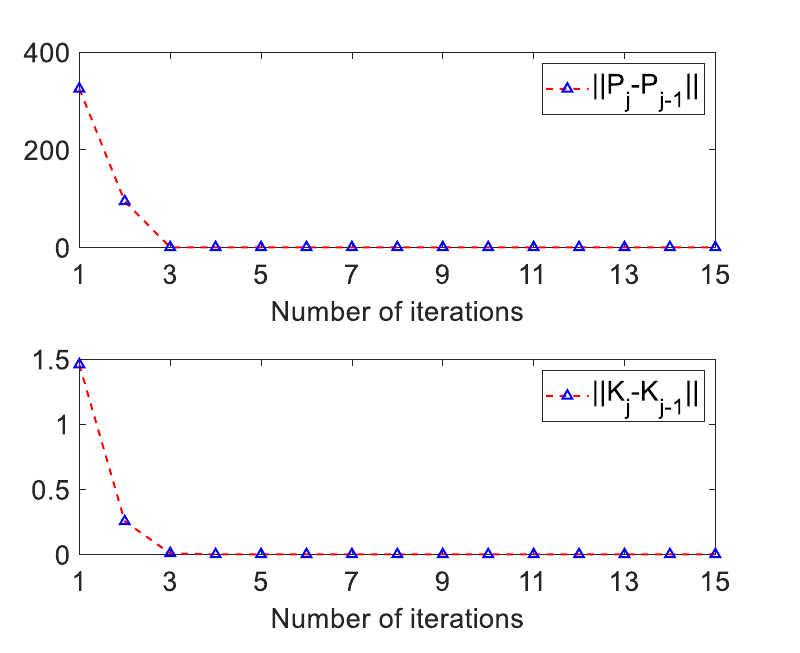}
 
    \caption{Convergence of matrix $P$ and feedback gain $K$.}
    \label{Fig:ConvADP}
\end{figure}

The authority allocation strategy is implemented to combine the control input from the driver and the controller. During simulation, the driver's input and the controller's input will be collected and their product will be integrated within time interval $\Delta t=5s$ to indicate the level of cooperativeness (\ref{Equ:Coop}). The automation authority will be set higher if the cooperation index is positive. Positive gains $\kappa_1$ and $\kappa_2$ are selected for different experimental scenarios such that the variation of authority will neither be too large or too small. In addition, it is worth noting that the system model is established only for simulation requirements and the accurate parameters will not be used in the controller design process since ADP is a model-free method. Before the actual control simulation starts, the optimal control problem is settled in advance using ADP algorithm. A test run will be conducted to gather the necessary data for iteration. The weight matrices are set and the iteration proceeds following Algorithm \ref{Algorithm1}. For the nonlinear part in CNF control, the parameters are $\phi=1\times10^{-4}$ and $\gamma=1$. In the test run, the matrix $P$ and the feedback gain $K$ are able to converge to optimal values, which can be seen from Fig. \ref{Fig:ConvADP}. Next, parameters including $B, U, X$ and $L$ can also be determined, which will be applied together with $P$ and $K$ to form the optimal control policy. Then, the formal operation begins based on the derived control policy and the self-triggered rule will take action to save computing resources. During this process, the system will compute the ET threshold $e_T$ and the next triggering instant $\tau_{k+1}$ in real time depending on the sampled system state $x_{ek}$ according to (\ref{Equ:ET}) and (\ref{Equ:ST}).

Driving conditions that exist in the real world including S-turn, U-turn, lane change, etc., can all be expressed in the form of road curvature profiles. Thus, the path tracking control method should be tested on roads with varying curvature to represent different driving scenarios. Two different cases are adopted for the simulation, including a quarter turn to the right and a complex loop trajectory tracking environment. Case I possesses a simple turning condition and Case II possesses a loop with frequent curvature changes. In order to clearly evaluate the lane keeping performance, an index $J_{rms}$ is utilized considering the root mean square value of the lateral deviation $y_c$, which is denoted as follows:
\begin{equation}
    J_{rms}=\sqrt{\frac{1}{T}\int_{0}^{T}{y_c}^2dt}
\end{equation}
where $T$ represents the total simulation time and smaller $J_{rms}$ indicates better lane keeping performance. The road conditions and the comparative results will be described in detail in the following.

\subsection{Case I: Quarter Turn to the Right}
\begin{figure}[!t]
    \centering
    \includegraphics[width=1\linewidth]{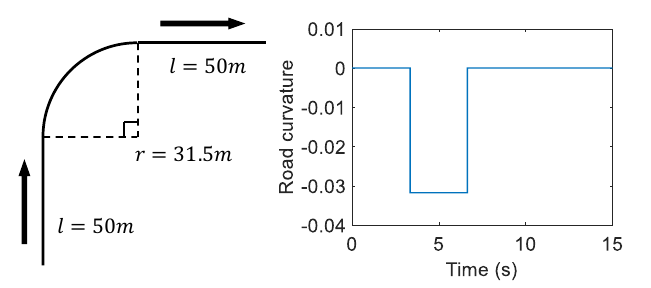}
    \caption{Road conditions for Case I.}
    \label{Fig:C1RdCond}
\end{figure}

In this quarter turn simulation, the vehicle enters from from the south and leaves from the east after passing a rightward road with radius of curvature $1/\rho=31.5m$, as illustrated in Fig. \ref{Fig:C1RdCond}. The total simulation time is 15 seconds, which is enough for the vehicle to finish the whole process. The parameter for calculating the cooperativeness level is $\kappa_1=5$. Three basic conditions are utilized to compare with the proposed method in this article:
\begin{enumerate}
    \item Controller with no CNF.
    \item Controller with no authority allocation (AA) strategy, i.e., the control authority of the driver and the automatic controller remain constant. Results are obtained for three different authority levels such that $\sigma=0.3, 0.5$ and $0.7$.
    \item Controller with no self-triggered rule. In this condition, the system becomes time-triggered and the triggering times are the main concern rather than the tracking performance.
\end{enumerate}

\begin{figure}[!t]
    \centering
    \begin{minipage}{\linewidth}
 	\centering
	\includegraphics[width=0.9\linewidth]{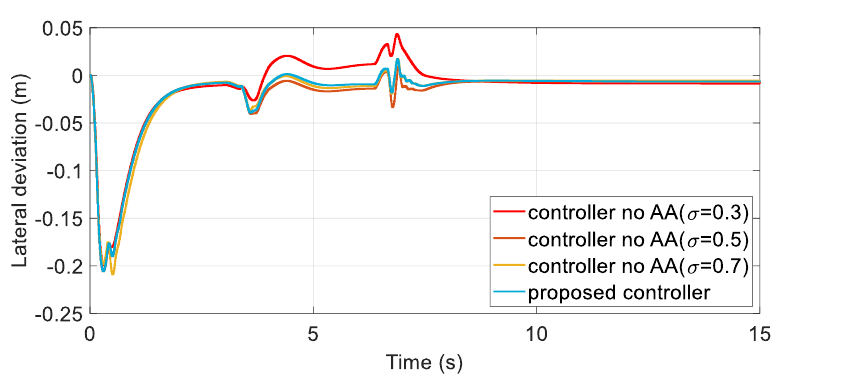}
    \end{minipage}
 
    \begin{minipage}{\linewidth}
	\centering
	\includegraphics[width=0.9\linewidth]{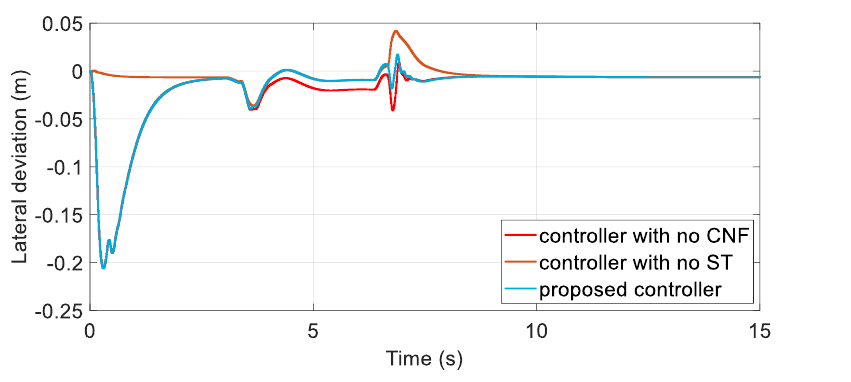}
    \end{minipage}
    \caption{Vehicle lane-keeping error in Case I.}
    \label{Fig:C1Lateral}
\end{figure}

\begin{figure}[!t]
    \centering
    \includegraphics[width=0.9\linewidth]{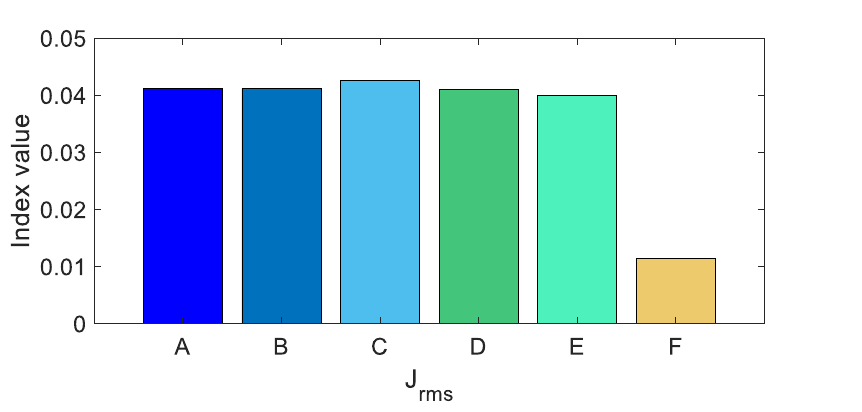}
    \caption{Performance indices in Case I. Starting from the left: A. Controller no AA($\sigma=0.3$). B. Controller no AA($\sigma=0.5$). C. Controller no AA($\sigma=0.7$). D. Controller with no CNF. E. Proposed controller. F. Controller with no ST.}
    \label{Fig:C1RMS}
\end{figure}

\begin{figure}[!t]
    \centering
    \includegraphics[width=0.9\linewidth]{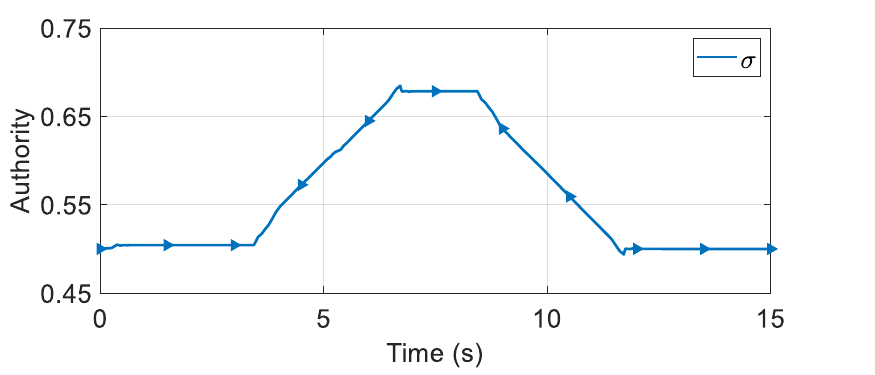}
    \caption{Automation authority level in Case I.}
    \label{Fig:C1Auth}
\end{figure}

Fig. \ref{Fig:C1Lateral} and Fig. \ref{Fig:C1RMS} show the lateral lane-keeping error and performance indices under several different circumstances in Case I. It can be seen that although the lateral deviation all converge to a small neighbourhood near the origin, the proposed method with CNF and authority allocation performs relatively better than others, especially in curved road sections. Compared with time-triggered control, namely the controller with no ST, the relatively large deviation of approximately -0.2$\mathrm{m}$ at the beginning of the simulation is mainly caused by the conflict between the initial state settings in the system and the ST mechanism, which is acceptable if it does not exceed a certain range. With CNF control, the transient performance are improved and the lateral error is reduced. Fig. \ref{Fig:C1Auth} illustrates the variation of automation control authority level over time. The level of automation authority is increased in well-cooperative situations, allowing the system to dominate the driving process to relief the driver's burden while reducing the tracking error.

\begin{figure}[!t]
    \centering
    \includegraphics[width=0.9\linewidth]{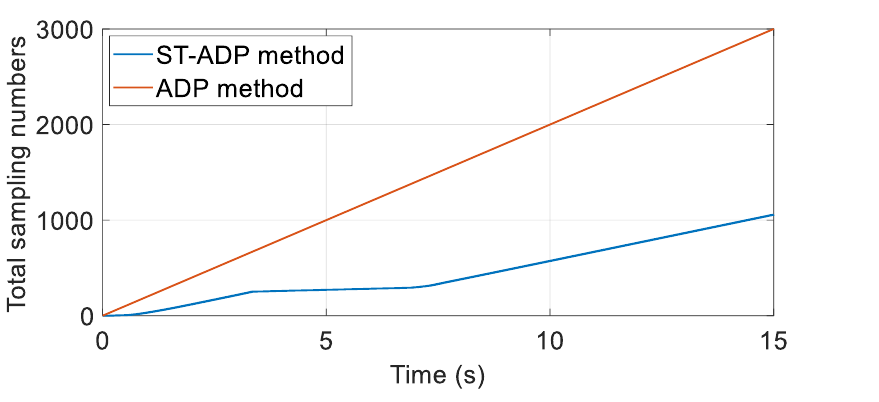}
    \caption{Comparison of triggering times in Case I.}
    \label{Fig:C1ST}
\end{figure}

Fig. \ref{Fig:C1ST} compares the evolution of triggering times in STC and in conventional time-triggered control. With the ST rule, the triggering interval can stabilize at some time intervals around $0.01s$, which is larger than the time-triggering interval $0.005s$ set in advance in the system. More specifically, the triggering rate is further reduced on particular occasions around $t=5s$. In total, the time-triggered control uses 3000 triggers while ST only needs 1057 triggers, which is a $64.77\%$ reduction of data communication and computational cost. The effectiveness of ST is validated.

\subsection{Case II: Complex Loop Environment}
\begin{figure}[!t]
    \centering
    \includegraphics[width=1\linewidth]{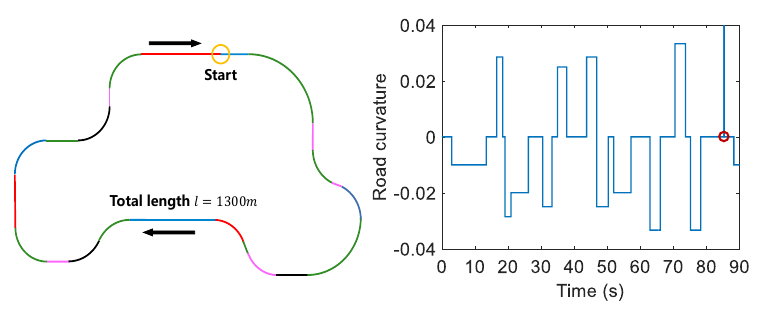}
    \caption{Road conditions for Case II.}
    \label{Fig:C2RdCond}
\end{figure}

Apart from the relatively simple environment, a complex loop trajectory tracking environment is adopted for more comprehensive tests. In this loop line, the vehicle departs form the starting point, passes through various road sections with different curvatures and finally returns to the starting point after finishing one lap, with the route map and the variation of road curvature illustrated in Fig. \ref{Fig:C2RdCond} where the red spot marks the return to the starting point. The entire length of the route is 1300$\mathrm{m}$ and the total simulation time is 90 seconds, which enables the vehicle to complete approximately one lap. The parameter for calculating the cooperativeness level is $\kappa_2=5$. Identical with Case I, three basic conditions are utilized to compare with the proposed method in this article: 
\begin{enumerate}
    \item Controller with no CNF.
    \item Controller with no authority allocation (AA) strategy, i.e., the control authority of the driver and the automatic controller remain constant. Results are obtained for three different authority levels such that $\sigma=0.3, 0.5$ and $0.7$.
    \item Controller with no self-triggered rule. In this condition, the system becomes time-triggered and the triggering times are the main concern rather than the tracking performance.
\end{enumerate}

\begin{figure}[!t]
    \centering
    \begin{minipage}{\linewidth}
 	\centering
	\includegraphics[width=0.9\linewidth]{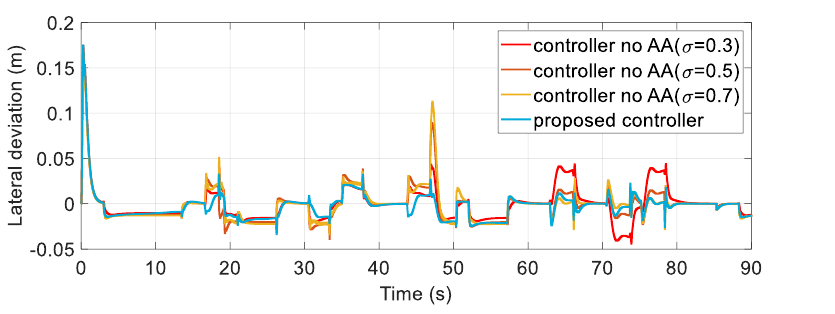}
    \end{minipage}
 
    \begin{minipage}{\linewidth}
	\centering
	\includegraphics[width=0.9\linewidth]{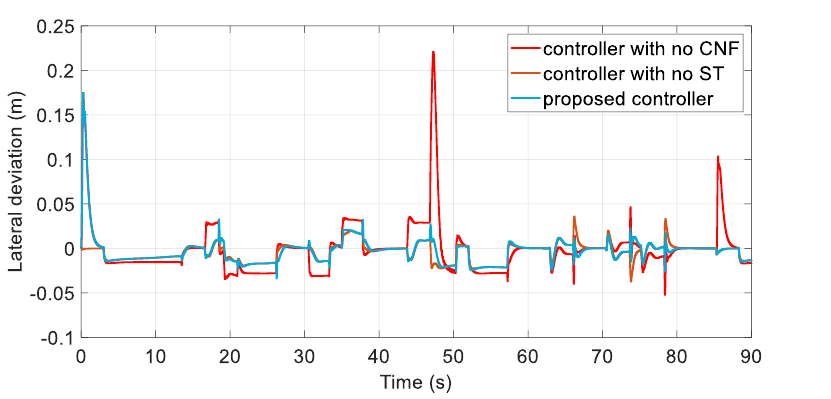}
    \end{minipage}
    \caption{Vehicle lane-keeping error in Case II.}
    \label{Fig:C2Lateral}
\end{figure}

\begin{figure}[!t]
    \centering
    \includegraphics[width=0.9\linewidth]{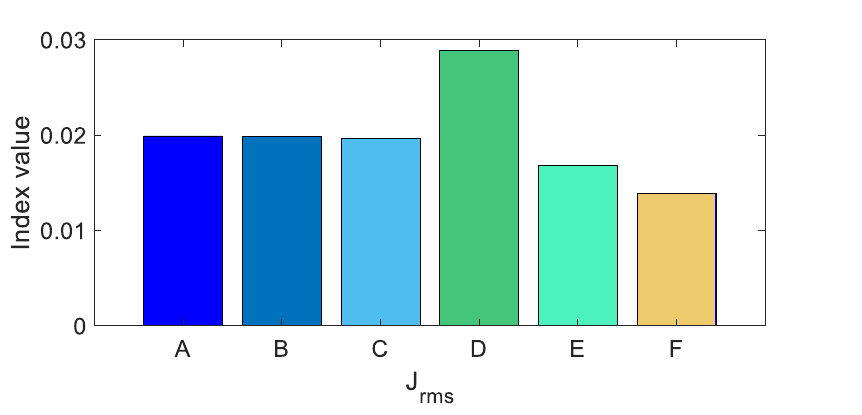}
    \caption{Performance indices in Case II. Starting from the left: A. Controller no AA($\sigma=0.3$). B. Controller no AA($\sigma=0.5$). C. Controller no AA($\sigma=0.7$). D. Controller with no CNF. E. Proposed controller. F. Controller with no ST.}
    \label{Fig:C2RMS}
\end{figure}

\begin{figure}[!t]
    \centering
    \includegraphics[width=0.9\linewidth]{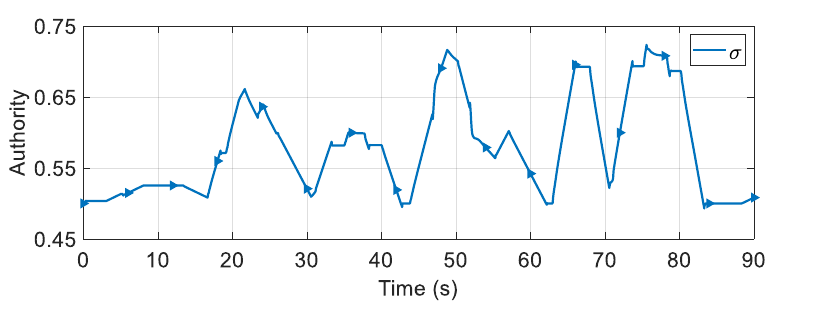}
    \caption{Automation authority level in Case II.}
    \label{Fig:C2Auth}
\end{figure}

Fig. \ref{Fig:C2Lateral} and Fig. \ref{Fig:C2RMS} show the lateral lane-keeping error and performance indices under several different circumstances in Case II. Among these curves, it can be seen that the proposed method with CNF and authority allocation has the best overall performance. For the controller with no CNF, the lane-keeping error is larger in the first half of the test and some large overshoots occur in the second half when the road curvature changes constantly. Notice that the large deviation appears at the beginning due to the conflict between the initial state settings and the ST mechanism compared with time-triggered control. It's clear that the transient performance are improved and the lateral error is reduced with CNF control. Fig. \ref{Fig:C2Auth} illustrates the variation of automation control authority level overtime. The level of automation authority changes with the actual situation, which is more flexible for the driver. The proposed controller performs better than the controller with no AA because of its smaller deviation and less overshoots.

\begin{figure}[!t]
    \centering
    \includegraphics[width=0.9\linewidth]{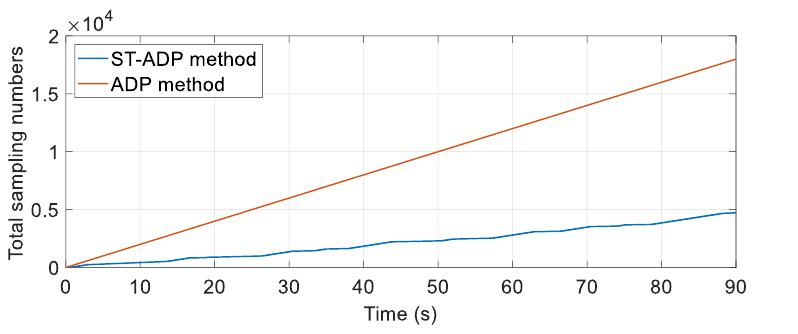}
    \caption{Comparison of triggering times in Case II.}
    \label{Fig:C2ST}
\end{figure}

Fig. \ref{Fig:C2ST} compares the evolution of triggering times in STC and in conventional time-triggered control. The time triggering interval is also set to be a constant $0.005s$ in the system. With the ST rule, the triggering times is 4724, which is reduced by $73.76\%$ compared with 18000 sampling numbers in time-triggered control, thus saving the computing resources. From the above simulation results, the effectiveness of the proposed controller in the shared control framework has been examined.

\section{Conclusion}
This paper has studied the CNF control method with ST-ADP algorithm in a shared steering control framework. A 2-DOF vehicle model and a two-point preview driver model are first established for mathematical analysis. The overall control input is obtained by combining the driver's input and the controller's input through an adaptive authority allocation strategy based on the level of cooperativeness. The CNF controller is designed for refining the transient performance. Using the ADP algorithm, the optimal control problem can be solved through iteration without knowing the accurate system parameters. After the control policy is obtained, the self-triggered rule enables the system to update in discrete times to save computing resources. Mathematical analysis and proofs have been presented and Carsim-Simulink simulations are carried out to verify the effectiveness of the proposed mechanism. In future work, we will focus on the impact of data quality on the computational results to make the ADP algorithm more robust. We will also test more control methods for comparison and carry out the experiment on a hardware driving simulator or a real passenger car.

\vspace{20pt}

\section{Biography Section}
 
\vspace{-30pt}


\begin{IEEEbiography}[{\includegraphics[width=1in,height=1.25in,clip,keepaspectratio]{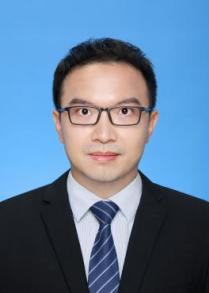}}]{Chuan Hu}
\\ is now a tenure-track Associate Professor at the School of Mechanical Engineering, Shanghai Jiao Tong University, Shanghai, China, starting from July, 2022. Before that, he was an Assistant Professor at the Department of Mechanical Engineering, University of Alaska Fairbanks, Fairbanks, AK, USA from June, 2020 to June, 2022. He was a Postdoctoral Fellow at Department of Mechanical Engineering, University of Texas at Austin, Austin, USA, from August, 2018 to June, 2020, and a Postdoctoral Fellow in the Department of Systems Design Engineering, University of Waterloo, Waterloo, Canada from July, 2017 to July, 2018. He received the Ph.D. degree in Mechanical Engineering, McMaster University, Hamilton, Canada in 2017, the M.S. degree in Vehicle Operation Engineering from the China Academy of Railway Sciences, Beijing, in 2013, and the B.S. degree in Automotive Engineering from Tsinghua University, Beijing, China, in 2010. His research interest includes the perception, decision-making, path planning, and motion control of Intelligent and Connected Vehicles (ICVs), Autonomous Driving (AD), eco-driving, human-machine trust and cooperation, shared control, and machine-learning applications in ICVs. He has published more than 60 papers in these research areas. He is currently an Associate Editor for several leading IEEE Trans. journals in this filed: IEEE Trans. on Neural Networks and Learning Systems, IEEE Trans. on Vehicular Technology, IEEE Trans. on Transportation Electrification, IEEE Trans. on Intelligent Transportation Systems and IEEE Trans. on Intelligent Vehicles. 
\end{IEEEbiography}
\vspace{11pt}

\begin{IEEEbiography}[{\includegraphics[width=1in,height=1.25in,clip,keepaspectratio]{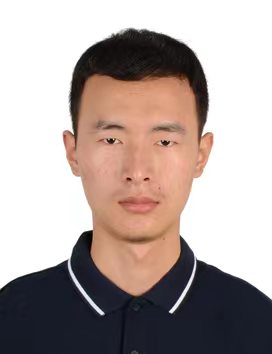}}]{Sicheng Ge}
\\ received the B.E. degree in mechanical engineering from Shanghai Jiao Tong University, Shanghai,
China in 2023. He is currently working towards
the Master degree in mechanical engineering at
the School of Mechanical Engineering, Shanghai
Jiao Tong University, Shanghai, China. His research
interest includes human-machine interaction, adaptive authority allocation policy and optimal control based on reinforcement learning for human-vehicle cooperative system.

\end{IEEEbiography}
\vspace{11pt}

\begin{IEEEbiography}[{\includegraphics[width=1in,height=1.25in,clip,keepaspectratio]{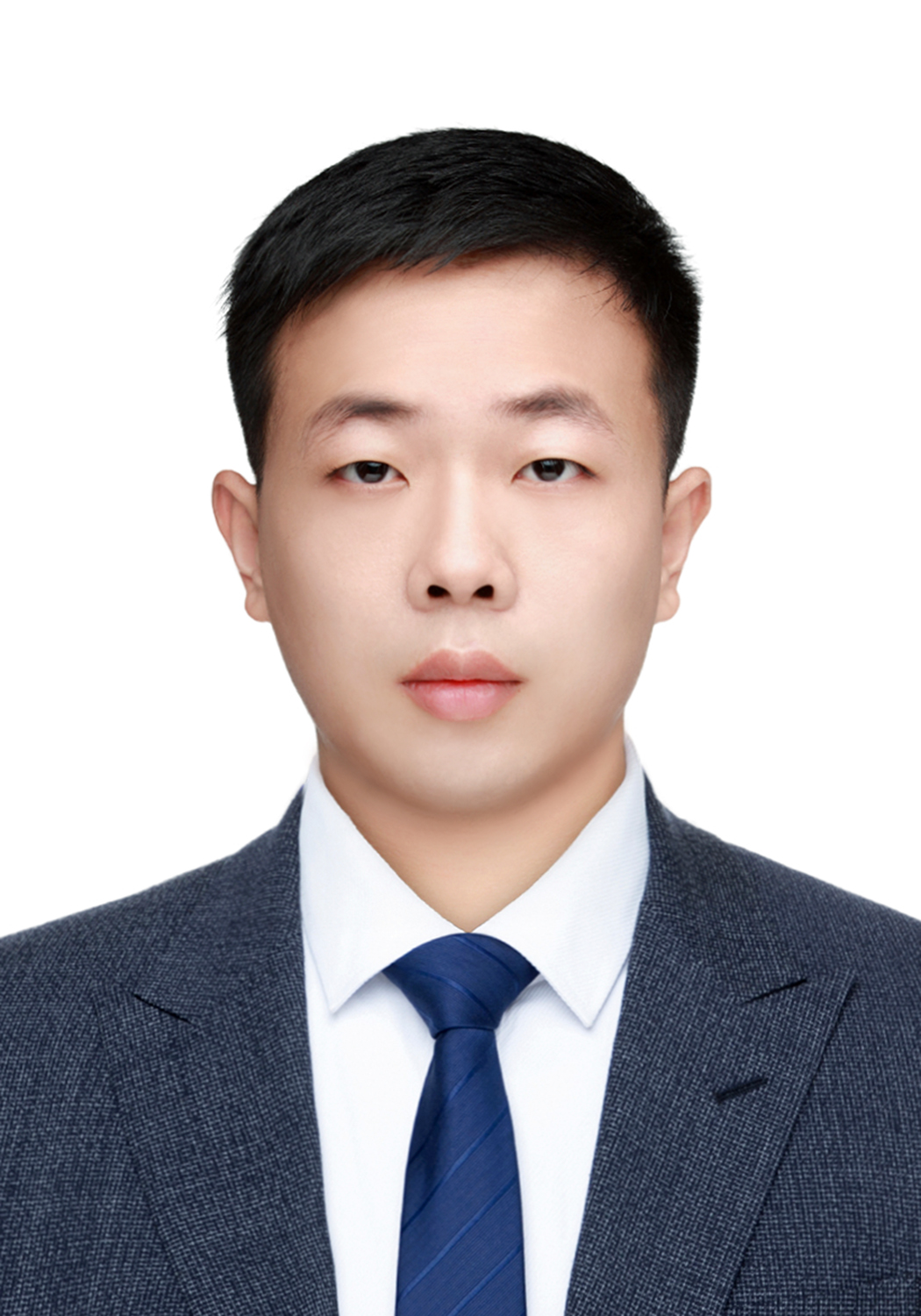}}]{Yingkui Shi}
\\ received the B.E. degree in mechanical engineering from Shanghai Jiao Tong University, Shanghai,
China in 2022. He is currently working towards
the Master degree in mechanical engineering at
the School of Mechanical Engineering, Shanghai
Jiao Tong University, Shanghai, China. His research
interest includes trust dynamics modeling, human-machine interaction and shared control algorithm based on game theory and reinforcement learning in human-vehicle cooperation areas.
\end{IEEEbiography}
\vspace{11pt}

\begin{IEEEbiography}[{\includegraphics[width=1in,height=1.25in,clip,keepaspectratio]{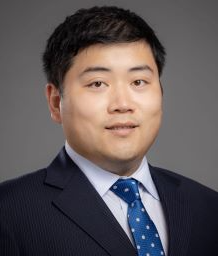}}]{Weinan Gao}
 (Senior Member, IEEE) received the B.Sc. degree in automation 
and the M.Sc. degree in control theory and control engineering from Northeastern University, Shenyang, China, in 2011 and 2013, respectively, and the Ph.D. degree in electrical engineering from New York University, Brooklyn, NY, USA, in 2017.
He is a Professor with the State Key Laboratory of Synthetical Automation for Process Industries at Northeastern University, Shenyang, China. Previously, he was an Assistant Professor of Mechanical and Civil Engineering at Florida Institute of Technology, Melbourne, FL, USA, an Assistant Professor of Electrical and Computer Engineering at Georgia Southern University, Statesboro, GA, USA, and a Visiting Professor of Mitsubishi Electric Research Laboratory (MERL), Cambridge, MA, USA.
His research interests include reinforcement learning, adaptive dynamic programming, optimal control, cooperative adaptive cruise control, intelligent transportation systems, sampled-data control systems, and output regulation theory.
He was a recipient of the Best Paper Award in IEEE International Conference on Real-Time Computing and Robotics (RCAR) in 2018 and the David Goodman Research Award at New York University in 2019. He was a recipient of the U.S.-NSF Engineering Research Initiation Award. He is a member of editorial board of \textit{Neural Computing and Applications} and a Technical Committee Member in IEEE Control Systems Society on Nonlinear Systems and Control and in IFAC TC 1.2 Adaptive and Learning Systems. He was an Associate Editor/a Guest Editor of IEEE/CAA \scshape{Journal of Automatica Sinica, IEEE Transactions on Neural Network and Learning Systems, IEEE Transactions on Circuits and Systems II: Express Briefs}, \textit{Neurocomputing}, \normalfont{and} \textit{Control Engineering Practice}.
\end{IEEEbiography}

\vspace{11pt}
\begin{IEEEbiography}[{\includegraphics[width=1in,height=1.25in,clip,keepaspectratio]{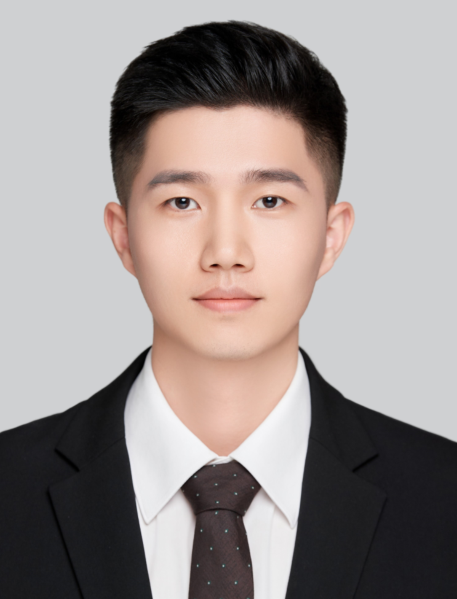}}]{Wenfeng Guo}
\\ received the B.E. degree in mechanical manufacture and automation and the Ph.D. degree in mechanical engineering from Hunan University, Changsha, China, in 2018 and 2023, respectively. He currently serves as a Post-Doctoral Researcher at the School of Vehicle and Mobility, Tsinghua University, Beijing, China. His research interests include driver-automation shared control, vehicle dynamics control, and driver assistance systems for intelligent vehicles. 
\end{IEEEbiography}

\begin{IEEEbiography}[{\includegraphics[width=1in,height=1.25in,clip,keepaspectratio]{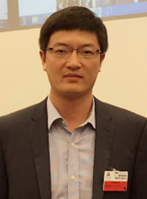}}]{Xi Zhang}
 (Senior Member, IEEE) received the B.Sc. degree in applied mathematics and the B.E. degree in information and control engineering from Shanghai Jiao Tong University (SJTU), Shanghai, China in 2002, and the M.E. and Ph.D. degrees in power electronics and electric power drive from SJTU, in 2004 and 2007, respectively.
From September 2007 to July 2009, he held a Postdoctoral position in the Department of Electrical and Computer Engineering, tat he University of Michigan-Dearborn, Dearborn, MI, USA.
He is currently a Full Professor in the Institute of Intelligent \& Connected Automobile and National Engineering Lab for Automotive Electronics and Control Technology, SJTU. His research interests include intelligent \& connected automobile, intelligent transportation, power management strategies, power electronics devices, and electric motor control systems for alternative-fuel vehicles.
\end{IEEEbiography}

\vfill


\begin{thebibliography}{99}
\bibliographystyle{IEEEtran}
\bibitem{ref1}
Y. Xia, F. Pu, S. Li, and Y. Gao, “Lateral path tracking control of autonomous land vehicle based on ADRC and differential flatness,” \textit{IEEE Trans. Ind. Electron.}, vol. 63, no. 5, pp. 3091–3099, May 2016.
\bibitem{ref2}
A. Eidehall, J. Pohl, F. Gustafsson, and J. Ekmark, “Toward autonomous collision avoidance by steering,” \textit{IEEE Trans. Intell. Transp. Syst.}, vol. 8, no. 1, pp. 84–94, Mar. 2007.
\bibitem{ref3}
V. Cerone, M. Milanese, and D. Regruto, “Combined automatic lane-keeping and driver’s steering through a 2-DOF control strategy,” \textit{IEEE Trans. Control Syst. Technol.}, vol. 17, no. 1, pp. 135–142, Jan. 2009.
\bibitem{ref4}
M. Flad, L. Frhlich, and S. Hohmann, “Cooperative shared control driver assistance systems based on motion primitives and differential games,” \textit{IEEE Trans. Human-Mach. Syst.}, vol. 47, no. 5, pp. 711–722, Oct. 2017.
\bibitem{ref5}
A. Benloucif, A.-T. Nguyen, C. Sentouh, and J.-C. Popieul, “Cooperative trajectory planning for haptic shared control between driver and automation in highway driving,” \textit{IEEE Trans. Ind. Electron.}, vol. 66, no. 12, pp. 9846–9857, Dec. 2019.
\bibitem{ref6}
Z. Fang, J. Wang, Z. Wang, J. Liang, Y. Liu, and G. Yin, “A human-machine shared control framework considering time-varying driver characteristics,” \textit{IEEE Trans. Intell. Veh.}, vol. 8, no. 7, pp. 3826–3838, Jul. 2023.
\bibitem{ref7}
W. Guo et al., "Toward consumer acceptance of cooperative driving systems: A human-centered shared steering control approach within a hierarchical framework," \textit{IEEE Trans. Consum. Electron.}, vol. 70, no. 1, pp. 635-645, Feb. 2024.
\bibitem{ref8}
W. Schwarting, J. Alonso-Mora, L. Paull, S. Karaman and D. Rus, “Safe nonlinear trajectory generation for parallel autonomy with a dynamic vehicle model," \textit{IEEE Trans. Intell. Transp. Syst.}, vol. 19, no. 9, pp. 2994-3008, Sept. 2018.
\bibitem{ref9}
X. Li, Y. Wang, C. Su, X. Gong, J. Huang, and D. Yang, “Adaptive authority allocation approach for shared steering control system,” \textit{IEEE Trans. Intell. Transp. Syst.}, vol. 23, no. 10, pp. 19428–19439, Oct. 2022.
\bibitem{ref10}
A.-T. Nguyen, C. Sentouh, and J.-C. Popieul, “Sensor reduction for driver-automation shared steering control via an adaptive authority allocation strategy,” \textit{IEEE/ASME Trans. Mechatron.}, vol. 23, no. 1, pp. 5–16, Feb. 2018.
\bibitem{ref11}
C. Sentouh, A.-T. Nguyen, M. A. Benloucif and J.-C. Popieul, “Driver-automation cooperation oriented approach for shared control of lane keeping assist systems," \textit{IEEE Trans. Control Syst. Technol.}, vol. 27, no. 5, pp. 1962-1978, Sept. 2019.
\bibitem{ref12}
C. Guo, C. Sentouh, J.-C. Popieul and J.-B. Haué, “Predictive shared steering control for driver override in automated driving: A simulator study," \textit{Transportation Research Part F: Traffic Psychology and Behaviour}, vol. 61, pp. 326-336, Feb. 2019.
\bibitem{ref13}
Z. Yan, K. Yang, Z. Wang, B. Yang, T. Kaizuka and K. Nakano, “Intention-based lane changing and lane keeping haptic guidance steering system," \textit{IEEE Trans. Intell. Veh.}, vol. 6, no. 4, pp. 622-633, Dec. 2021.
\bibitem{ref14}
A.-T. Nguyen, J. J. Rath, C. Lv, T.-M. Guerra and J. Lauber, “Human-machine shared driving control for semi-autonomous vehicles using level of cooperativeness,” \textit{Sensors}, vol. 21, no. 14, p. 4647, Jul. 2021.
\bibitem{ref15}
R. Potluri and A. K. Singh, “Path-tracking control of an autonomous 4WS4WD electric vehicle using its natural feedback loops,” \textit{IEEE Trans. Control Syst. Technol.}, vol. 23, no. 5, pp. 2053–2062, Sep. 2015.
\bibitem{ref16}
T. Yuan and R. Zhao, “LQR-MPC-based trajectory-tracking controller of autonomous vehicle subject to coupling effects and driving state uncertainties,” \textit{Sensors}, vol. 22, no. 15, p. 5556, Jul. 2022.
\bibitem{ref17}
H. Li, K. Liu, B. Yang, L. Zhang and Y. Yan, “Path tracking of autonomous vehicle based on NMPC with pre-steering," \textit{IEEE Trans. Consum. Electron.}, vol. 70, no. 1, pp. 966-979, Feb. 2024.
\bibitem{ref18}
L. Zhai, C. Wang, Y. Hou, and C. Liu, “MPC-based integrated control of trajectory tracking and handling stability for intelligent driving vehicle driven by four hub motor,” \textit{IEEE Trans. Veh. Technol.}, vol. 71, no. 3, pp. 2668–2680, Mar. 2022.
\bibitem{ref19}
D. Liu, S. Xue, B. Zhao, B. Luo, and Q. Wei, “Adaptive dynamic programming for control: A survey and recent advances,” \textit{IEEE Trans. Syst., Man, Cybern., Syst.}, vol. 51, no. 1, pp. 142–160, Jan. 2021.
\bibitem{ref20}
Y. Jiang and Z. Jiang, “Computational adaptive optimal control for continuous-time linear systems with completely unknown dynamics,” \textit{Automatica}, vol. 48, no. 10, pp. 2699–2704, 2012.
\bibitem{ref21}
M. Huang, W. Gao, Y. Wang, and Z.-P. Jiang, “Data-driven shared steering control of semi-autonomous vehicles,” \textit{IEEE Trans. Human-Mach. Syst.}, vol. 49, no. 4, pp. 350–361, Aug. 2019.
\bibitem{ref22}
W. Sun, X. Wang and C. Zhang, "A model-free control strategy for vehicle lateral stability with adaptive dynamic programming," \textit{IEEE Trans. Ind. Electron.}, vol. 67, no. 12, pp. 10693-10701, Dec. 2020.
\bibitem{ref23}
H. Dong, X. Zhao, and B. Luo, “Optimal tracking control for uncertain nonlinear systems with prescribed performance via critic-only ADP,” \textit{IEEE Trans. Syst., Man, Cybern., Syst.}, vol. 52, no. 1, pp. 561–573, Jan. 2022.
\bibitem{ref24}
H. Wang, Z. Zuo, Y. Wang and H. Yang, "Composite nonlinear path-following control for unmanned ground vehicles with anti-windup ESO," \textit{IEEE Trans. Syst., Man, Cybern., Syst.}, vol. 52, no. 9, pp. 5865-5876, Sept. 2022.
\bibitem{ref25}
R. Wang, C. Hu, F. Yan, and M. Chadli, “Composite nonlinear feedback control for path following of four-wheel independently actuated autonomous ground vehicles,” \textit{IEEE Trans. Intell. Transp. Syst.}, vol. 17, no. 7, pp. 2063–2074, Jul. 2016.
\bibitem{ref26}
B. M. Chen, T. H. Lee, K. Peng, and V. Venkataramanan, “Composite nonlinear feedback control for linear systems with input saturation: Theory and an application,” \textit{IEEE Trans. Autom. Control}, vol. 48, no. 3, pp. 427–439, Mar. 2003.
\bibitem{ref27}
L. Dong, X. Zhong, C. Sun, and H. He, “Event-triggered adaptive dynamic programming for continuous-time systems with control constraints,” \textit{IEEE Trans. Neural Netw. Learning Syst.}, vol. 28, no. 8, pp. 1941–1952, Aug. 2017.
\bibitem{ref28}
R. Song and L. Liu, “Event-triggered constrained robust control for partly-unknown nonlinear systems via ADP,” \textit{Neurocomputing}, vol. 404, pp. 294–303, Sep. 2020.
\bibitem{ref29}
F. Zhao, W. Gao, Z.-P. Jiang and T. Liu, "Event-triggered adaptive optimal control with output feedback: An adaptive dynamic programming approach," \textit{IEEE Trans. Neural Netw. Learning Syst.}, vol. 32, no. 11, pp. 5208-5221, Nov. 2021.
\bibitem{ref30}
Z. Ming, H. Zhang, Y. Yan and J. Sun, "Self-triggered adaptive dynamic programming for model-free nonlinear systems via generalized fuzzy hyperbolic model," \textit{IEEE Trans. Syst., Man, Cybern., Syst.}, vol. 53, no. 5, pp. 2792-2801, May. 2023. 
\bibitem{ref31}
K. Zhang, B. Zhou, W. X. Zheng and G.-R. Duan, "Event-triggered and self-triggered gain scheduled control of linear systems with input constraints," \textit{IEEE Trans. Syst., Man, Cybern., Syst.}, vol. 52, no. 10, pp. 6452-6463, Oct. 2022.
\bibitem{ref32}
R. Rajamani, \textit{Vehicle Dynamics and Control}. New York, NY, USA: Springer, 2011.
\bibitem{ref33}
W. Gao and Z. P. Jiang, “Adaptive dynamic programming and adaptive optimal output regulation of linear systems,” \textit{IEEE Trans. Autom. Control}, vol. 61, no. 12, pp. 4164–4169, Dec. 2016.
\bibitem{ref34}
B. M. Chen, T. H. Lee, K. Peng, and V. Venkataramanan, “Composite nonlinear feedback control for linear systems with input saturation: Theory and an application,” \textit{IEEE Trans. Autom. Control}, vol. 48, no. 3, pp. 427–439, Mar. 2003.
\bibitem{ref35}
H. Khalil, \textit{Nonlinear Systems}, 3rd ed. Upper Saddle River, NJ, USA: Prentice-Hall, 2002.
\bibitem{ref36}
D. Kleinman, “On an iterative technique for Riccati equation computations,” \textit{IEEE Trans. Autom. Control}, vol. 13, no. 2, pp. 114–115, Feb. 1968.

\end{thebibliography}
\end{document}